\documentclass[aps,prd,twocolumn,showpacs,groupedaddress,amsmath,amssymb,superscriptaddress,nofootinbib]{revtex4-2} %

\usepackage[dvipsnames]{xcolor}

\usepackage{graphicx}%
\usepackage{dcolumn}%
\usepackage{multirow} %
\usepackage{bm}%
\usepackage[colorlinks,linkcolor=blue,urlcolor=blue,citecolor=blue]{hyperref}

\usepackage{color}
\usepackage{soul}
\usepackage[utf8]{inputenc}
\usepackage[bottom]{footmisc}
\usepackage[T1]{fontenc}
\usepackage{mathptmx}
\usepackage{url}
\usepackage{relsize}

\usepackage{float}

\newcommand\unit[1]{{\rm #1}}

\newcommand\apjl{Astrophys. J. Lett.}

\newcommand\cqg{Class. Quantum Grav.}

\def\RIT{Center for Computational Relativity and Gravitation, Rochester Institute of Technology, Rochester, New York 14623, USA}
\def\GLA{SUPA, School of Physics \& Astronomy, University of Glasgow, Glasgow G12 8QQ, United Kingdom}

\begin{document}
	
\title{Efficient reanalysis of events from GWTC-3 with RIFT and asimov}

\author{D. Fernando}
\affiliation{\RIT}
\author{Richard O'Shaughnessy}
\affiliation{\RIT}
\author{Daniel Williams}
\affiliation{\GLA}

\begin{abstract}
Different waveform models can yield notably different conclusions about the 
properties of individual gravitational wave events. 
For instance, previous analyses using the SEOBNRv4PHM, IMRPhenomXPHM models, 
and NRSur7dq4 have led to varying results regarding event properties. 
This variability complicates the interpretation of the data and understanding 
of the astrophysical phenomena involved. 
There is an ongoing need to reassess candidate events with the best available interpretations and models. 
Current approaches lack efficiency or consistency, making it challenging to perform
large-scale reanalyses with updated models or improved techniques.  
It is imperative that investigations into waveform systematics be reproducible.  
Frameworks like asimov can facilitate large-scale reanalyses with consistent settings and high-quality results, 
and can reliably show how different waveform models affect the interpretation of gravitational wave events. 
This, in combination with other provided tools, allow for reanalysis of several events from the GWTC-3 catalog. 
We include access to full analysis settings that facilitate public use of GWOSC data on the Open Science Grid, particularly those conducted with the IMRPhenomPv2, SEOBNRv4PHM, SEOBNRv5PHM, and NRSur7dq4 waveform models. 
Our parameter inference results find similar conclusions to previously published work: for several events, all models
largely agree, but for a few exceptional events these models disagree substantially on the nature of the merging binary.

\end{abstract}

\maketitle

\section{Introduction}

Since the  discovery of GW150914\_095045 by the Advanced
LIGO \cite{2015CQGra..32g4001L} and Virgo \cite{gw-detectors-Virgo-original-preferred,2015CQGra..32b4001A} detectors,
now joined by KAGRA \cite{2021PTEP.2021eA101A},
the first four observing runs (referred to as O1, O2, 
O3, and O4) of the LIGO/Virgo/KAGRA network by the LVK Collaboration have identified the characteristic gravitational
wave signature of $O(\simeq 100)$ coalescing compact binaries \cite{LIGO-O3-O3b-catalog}.  
Independent
analyses of the public data have also contributed to the candidate event census; see, e.g., \cite{2023ApJ...946...59N,2020PhRvD.101h3030V,2023arXiv231206631W}
and references therein.
For each candidate event, the GW strain data can be compared to approximate models for the emitted gravitational radiation from a
coalescing (quasicircular) compact binary, to deduce the distribution of plausible source binary parameters consistent
with the data; see, e.g., \cite{LIGO-O3-O3b-catalog,gwastro-RIFT-Update} and references therein.

Lacking an efficient global solution for the two-body problem in general relativity, inferences about compact binary
parameters are performed with one or more of several approximate models for the outgoing radiation.  As a concrete
example, analysis of new events in  the GWTC-3 catalog paper employed two waveform models, known as SEOBNRv4PHM \cite{2018PhRvD..98h4028C,2020PhRvD.102d4055O} and
IMRPhenomXPHM  \cite{gwastro-mergers-IMRPhenomXP}, as did a subsequent reanalysis with a different inference strategy \cite{2023PhRvL.130q1403D}.  A recent study \cite{gwastro-mergers-TousifGWTC3} reanalyzed many of these same events with yet another waveform model,
NRSur7dq4 \cite{2019PhRvR...1c3015V}.  
As in previous analyses real events such as the GWTC-2 catalog  \cite{LIGO-O3-O3a-catalog}, these waveform models can
and do draw somewhat different conclusions about individual events' properties.    Both the GWTC-3 catalog and its
reanalysis identified several events with notably different properties \cite{LIGO-O3-O3a-catalog,gwastro-mergers-TousifGWTC3}.  More broadly, many investigations
repeatedly identify differences between even the best presently-available waveform models, such that the properties of
some astrophysically currently accessible events will be notably different when interpreted with different models 
\cite{gwastro-Systematics-Williamson2017,gwastro-systematics-ScottFeroz2019,LIGO-O3-O3bcatalog,LIGO-O3-GW190412,LIGO-O3-GW190521-implications,2023arXiv230318046R,gwastro-mergers-TousifGWTC3}.

Given the substantial and ongoing need to reassess candidate events with the best available interpretations, frameworks
that enable efficient large-scale reanalysis of events have become vital tools for gravitational wave astronomy.
In this work, we apply and refine one framework to manage analyses of many events with consistent settings: asimov \cite{williams2022asimov},
managing the RIFT parameter inference engine \cite{gwastro-PENR-RIFT,gwastro-RIFT-Update}.  Our full analysis settings and software are publicly available, including
worked exmaples on how to use public GWOSC data and the Open Science Grid to analyze some of these events \cite{2024settings}.
In an associated technical data releases and appendicies, we demonstrate how to use previous asimov settings, generate
asimov-compatible settings from existing public  analyses \cite{gwastro-mergers-TousifGWTC3}, and operate the end-to-end
production of our inference results.

As an application of our framework, we revisit several of the analyses presented in GWTC-3 with SEOBNRv4PHM:
specifically, the high-mass events reanalyzed with NRSur7dq4  \cite{gwastro-mergers-TousifGWTC3}.  Though
performed with RIFT, the previously published GWTC-3 analyses provided extremely few posterior samples, due to their use of
calibration marginalization implemented via  rejection sampling.  In this work, given the negligible impact of
calibration marginalization on astrophysical source properties, we perform comparable reanalyses to produce large,
high-quality sets of posterior samples without calibration marginalization.   
To demonstrate the utility of our framework for reproducible investigation of waveform systematics, we  also perform
these analyses with several other contemporary waveform models, including IMRPhenomPv2 \cite{gwastro-mergers-IMRPhenomP}, NRSur7dq4, SEOBNRv4PHM \cite{2018PhRvD..98h4028C,2020PhRvD.102d4055O}, and SEOBNRv5PHM \cite{2023arXiv230318046R,2023arXiv230318203M}.

This paper is organized as follows.  In Section \ref{sec:methods} we describe our parameter inference approach, the
asimov infrastructure, and the selected O3 events reanalyzed in this work.  In Section \ref{sec:results} we demonstrate
our reanalyses of these events.  We first describe selected anecdotal examples to highlight waveform systematics.  Then,
following previous studies, we present large-scale summary figures and tables, characterizing all the events used.

\section{Methods}
\label{sec:methods}

\subsection{Parameter inference with RIFT }

In a coalescing quasicircular orbit, a compact binary can be fully defined by 15 parameters. Among these, eight are the intrinsic parameters - signified by $\lambda$ - which encompass any factors describing the physical properties of the individual objects 
within the system - the binary's masses, spin magnitudes, spin angles, and azimuthl spin angles. 
Extrinsic parameters - signified by $\theta$ - consist of the other seven values necessary to describe its spacetime position and 
orientation relative to the detectors. 

RIFT interprets gravitational wave observations $d$ by comparing them 
with predicted gravitational wave signals $h(\bm{\lambda}, \bm\theta)$ in a two-step iterative process. 
In the first stage, RIFT has numerous workers concurrently calculate a marginal likelihood
\begin{equation}
 {\cal L}{({\bm \lambda})}\equiv\int  {\cal L}_{\rm full}(\bm{\lambda} ,\bm\theta )p(\bm\theta )d\bm\theta
\end{equation}
for many different values of $\bm{\lambda}$, where ${\cal L}_{\rm full}(\bm{\lambda} ,\theta ) $ is the likelihood of the gravitational wave 
signal in the multi-detector network; see \cite{gwastro-PE-AlternativeArchitectures,gwastro-PENR-RIFT} 
for further details.
In the second stage, RIFT will use accumulated marginal likelihood evaluations $(\bm{\lambda}_\alpha,{\cal L}_\alpha)$ 
to construct an approximation to ${\cal L}(\bm{\lambda})$ which is then used to infer the detector-frame posterior distribution
\begin{equation}
\label{eq:post}
p_{\rm post}=\frac{{\cal L}(\bm{\lambda} )p(\bm{\lambda})}{\int d\bm{\lambda} {\cal L}(\bm{\lambda} ) p(\bm{\lambda} )}.
\end{equation}
where the prior $p(\bm{\lambda})$ represents the prior distribution on intrinsic parameters.

RIFT inference is performed using the spin-weighted spherical harmonic waveforms  $h_{lm}(t)$ or $h_{lm}(f)$ \cite{gwastro-PE-AlternativeArchitectures,gwastro-PE-AlternativeArchitecturesROM,gwastro-PENR-RIFT,gwastro-PENR-RIFT-GPU}, usually computed from binary parameters
through the \text{lalsimulation} or \text{gwsignal} library.
In this work, we will  present RIFT inferences performed using several families of models.  For the SEOB family of
models, we employ 
SEOBNRv4PHM \cite{2018PhRvD..98h4028C,2020PhRvD.102d4055O} and  SEOBNRv5PHM
\cite{2023arXiv230318046R,2023arXiv230318203M}.  Among surrogate models calibrated to numerical relativity, we employ
the NRSur7dq4 model  \cite{2019PhRvR...1c3015V}.
While we will not use RIFT to analyze these events with IMRPhenomXPHM  \cite{gwastro-mergers-IMRPhenomXP}, we will
compare our results to  analyses performed by others, including analyses using this waveform.

\subsection{Asimov automation and standardization}
The asimov infrastructure provides a framework for managing suites of parameter inference calculations.  This framework
in particular provides not only a mechanism for storing event-specific and algorithm-specific settings, but also a
mechanism to implement these settings, launching and overseeing production analyses with a variety of parameter
inference tools.   Starting with the GWTC-3 and GWTC-2.1 papers, the LVK began performing its analyses using asimov; in
particular, the asimov configurations needed to perform these analyses are available \cite{williams2022asimov}.
In this work, we minimally adapt these settings with pipeline- and approximant-specific adjustments.  

The default analysis settings define a specific frequency range for each event, usually covering from $20\unit{Hz}$ up to some
maximum frequency $0.875f_{\rm PSD}$ where $f_{\rm PSD}$ is the largest frequency used in the original LVK PSD.

Other groups, however, have performed their own large-scale reanalyses using their own framework and metadata storage.
To facilitate comparison with specific sets of previous reanalyses \cite{gwastro-mergers-IMRPhenomXP} and \cite{gwastro-mergers-TousifGWTC3}, we implemented and provide
tools to convert from their metadata format to a standard asimov configuration file. 
Since not all of the metadata used for previous analyses is compatible with asimov, some differences are  expected
between the original analyses and our attempt at reproducing a comparable analysis.
For our analyses, we use asimov to get the PSDs by running Bayeswave for all events. However, we also provide tools to retrieve PSDs from previous publications and provide a demonstration on how this method also be used. 

\subsection{Selected O3 events and settings}
The  GWTC-3 paper, covering the second half of O3 (henceforth denoted as O3b), presented several analyses performed by
the LVK, including many events analyzed with SEOBNRv4PHM via RIFT.    The default design choices underlying these
analyses prioritized matching the configuration of corresponding analyses with another code, notably including the
effects of calibration marginalization.    Thus, after first  using RIFT to analyze events with SEOBNRv4PHM, these analyses were postprocessed with rejection
sampling, to account for the impact of calibration marginalization.   Previous studies have demonstrated that
calibration marginalization has little impact  on intrinsic or extrinsic parameters, modulo a modest effect in some
cases on the source sky location.   By contrast, the decimation performed by the O3b rejection sampling provides
surprisingly few samples to downstream users.  For this reason,  the reanalyses presented below will omit calibration
marginalization entirely.

Table \ref{table:event_info} lists the events we have selected for reanalysis.   In our demonstration, we chose to
reanalyze the same set of O3b events
recently reanalyzed with NRSur7dq4 \cite{gwastro-mergers-TousifGWTC3}.  The table also provides  metadata needed to initiate a
RIFT analysis: the specific event time and approximate mass range to
investigate.  By default, RIFT accesses this information from  a single file, nominally  produced by real search codes (a
``coincidence'' file commonly denoted \texttt{coinc.xml}), but alternatively generated from median search results.
For complete reproducibility of
pre-existing LVK analyses,
we  list meta-information enumerated by the event's internal gracedb characteristic event identifier along with a secondary
identifier identifying the characteristic event geocenter time and mass.    These identifiers sometimes differ from the
preferred released event, because of small but important differences between full inference and  the event time or other
parameters targeted by searches.  For public reanalyses, we have also
generated our own alternative characteristic information, disseminated along with our configuration files.

\begin{table}[htbp]
	\centering
	\begin{tabular}{|l|l|l|l|}
		\hline
		\textbf{Event} & \textbf{Superevent} & \textbf{GID} & \textbf{GPS Time}\\
		\hline
		GW191109\_010717 & S191109d & G354219 & 1257296855.216458\\
		GW191222\_033537 & S191222n & G358088 & 1261020955.123615\\
		GW191230\_180458 & S191230an & G358883 & 1261764316.406738\\
		GW200112\_155838 & S200112r & G359994 & 1262879936.090936\\
		GW200128\_022011 & S200128d & G361407 & 1264213229.901367\\
		GW200129\_065458 & S200129m & G361581 & 1264316116.433214\\
		GW200208\_130117 & S200208q & G363349 & 1265202095.949707\\
		GW200209\_085452 & S200209ab & G363458 & 1265273710.17476\\
		GW200216\_220804 & S200216br & G364262 & 1265926102.879405\\
		GW200219\_094415 & S200219ac & G364600 & 1266140673.196691\\
		GW200220\_061928 & S200220ad & G364748 & 1266214786.632\\
		GW200220\_124850 & S200220aw & G364783 & 1266238148.15\\
		GW200224\_222234 & S200224ca & G365380 & 1266618172.401773\\
		GW200302\_015811 & S200302c & G366190 & 1267149509.516065\\
		GW200311\_115853 & S200311bg & G367788 & 1267963151.398\\
		\hline
	\end{tabular}
	\caption{Event Information}
	\label{table:event_info}
\end{table}

In this work, we will perform inference on either the publicly-released GWTC-3 data from GWOSC \cite{2021SoftX..1300658A}, or
(to facilitate use of settings adopted in prior work) bit-equivalent internal data.   Following the analyses used for
LVK publications, we use GWOSC-released deglitched data when appropriate.  Unless otherwise noted, we generate independent PSD
estimates using Bayeswave \cite{2015CQGra..32m5012C,2015PhRvD..91h4034L,2019PhRvD.100j4004C}.
However, while as a second alternative and for comparison, when alternatively gathered PSDs from existing analyses of the
GWTC-3 catalog \cite{LIGO-O3-O3b-catalog}; examples of this configuration are also disseminated in our configuraiton
repository \cite{2024settings}.

\section{Results}
\label{sec:results}

\subsection{Summary statistics}
\begin{figure*}
	\includegraphics[width=\textwidth]{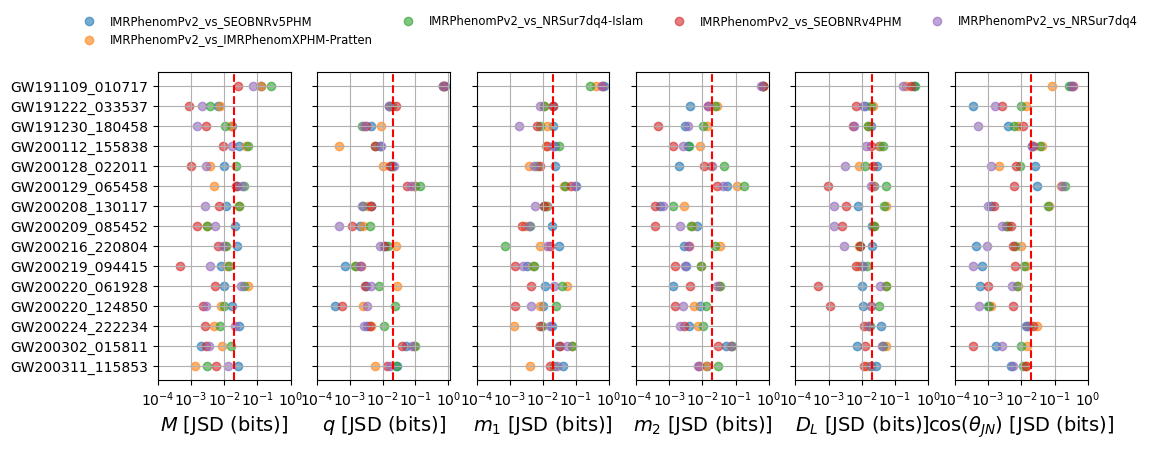}
	\vspace{0.5cm}
	\includegraphics[width=\textwidth]{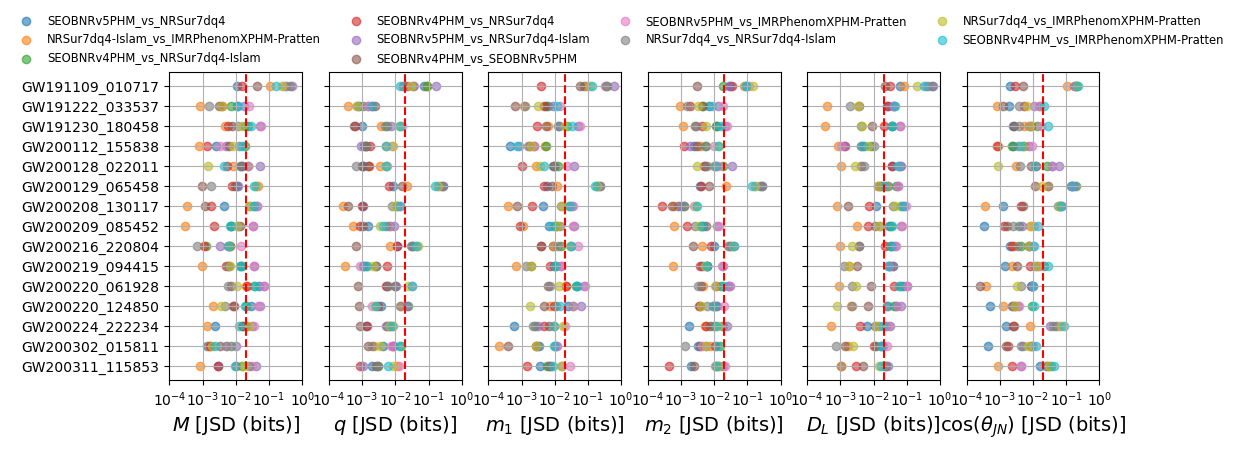}
	\caption{Jenson-Shannon (JS)  divergence between the one-dimensional marginalized posteriors of the source-frame total
	  mass $M$, the mass ratio $q$, source frame component masses $m_i$, luminosity distance $D_L$, and inclination angle
	  $\theta_{JN}$.  Colors indicate the two different waveform models used in the comparison.  The vertical dashed red
	  line corresponds to a JS divergence of $0.02$, above which differences between one-dimensional marginal distributions
	  are easily identified by eye.   
	\label{fig:js_batch1}
	}
\end{figure*}

\begin{figure*}
	\includegraphics[width=\textwidth]{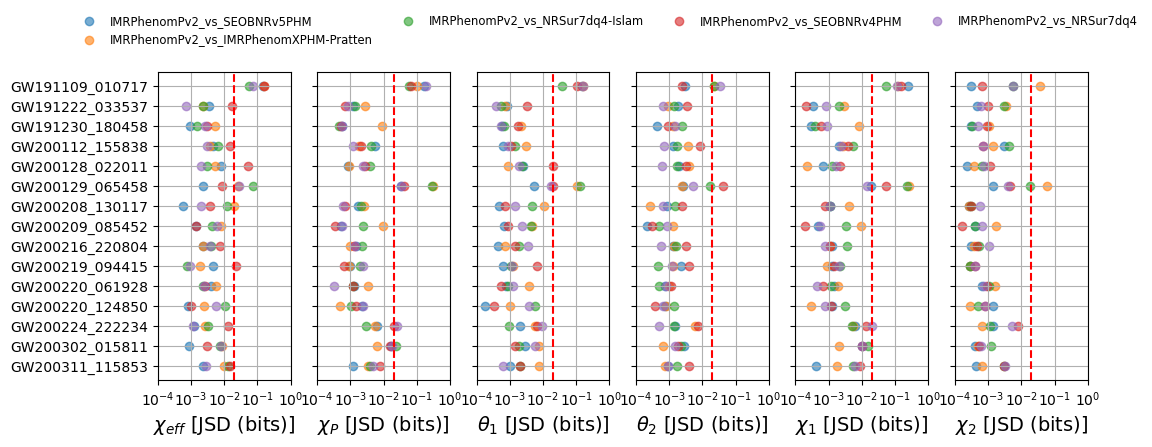}
	\includegraphics[width=\textwidth]{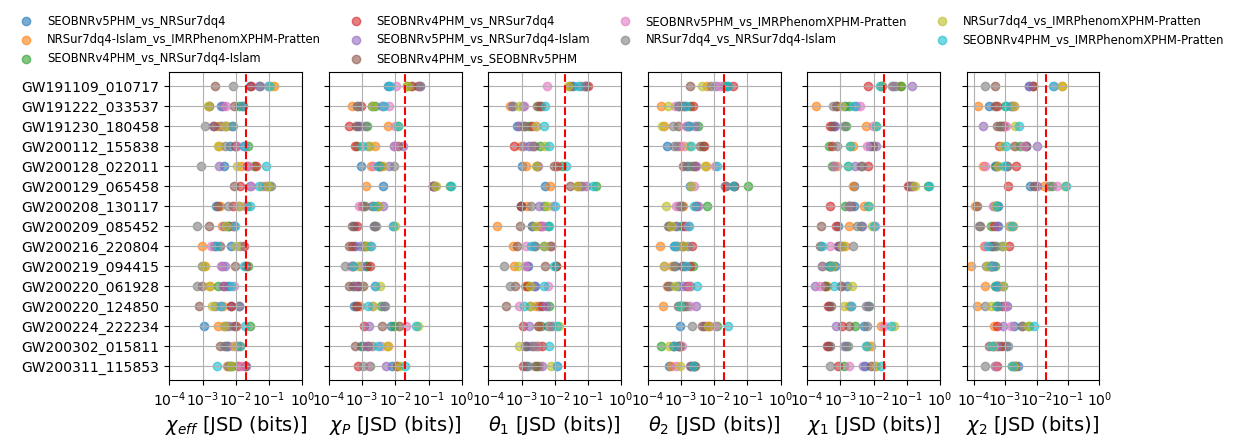}
	\caption{The Jenson-Shannon (JS)  divergence between
	the one-dimensional marginalized posteriors of the
	  effective inspiral spin parameter $\chi_{eff}$, 
	  spin precession parameter $\chi_p$, spin angles $\theta_1$ and $\theta_2$, and spin
	  magnitudes $\chi_1$ and $\chi_2$.   Colors indicate the two different waveform models used in the comparison.  The vertical dashed red
	  line corresponds to a JS divergence of $0.02$, above which differences between one-dimensional marginal distributions
	  are easily identified by eye.   
	\label{fig:js_batch2}
	}
\end{figure*}

Following \cite{gwastro-mergers-TousifGWTC3}, Figures \ref{fig:js_batch1} and \ref{fig:js_batch2} illustrate a simple
summary statistic -- the Jensen-Shannon (JS) divergence --  quantifying differences between marginal one-dimensional
posterior distributions inferred for each event.    The  vertical dashed red line corresponds to a JS divergence of $0.02$, above which differences between one-dimensional marginal distributions
  are easily identified by eye. 
The top panels of Figures \ref{fig:js_batch1} and \ref{fig:js_batch2} show comparisons between inferences with three state-of-the-art
waveform models compared to inferences derived with an older model (IMRPhenomPv2) known to omit important physics and
higher-order modes.  Except for a handful of notable exceptions (GW191109, GW200302, GW200129), most analyses with
IMRPhenomPv2 are broadly consistent with analyses done with more sophisticated waveforms.  We demonstrate how this
counterintuitive result arises using concrete examples in the next section.
By contrast,  the bottom panels of Figures \ref{fig:js_batch1} and \ref{fig:js_batch2} show familiar differences when
comparing results between different state-of-the-art waveform models.  These comparisons suggest that almost every event
has some parameter in which notable waveform differences occur.  Likewise, these comparisons suggest that every pair of
waveforms (including SEOBNRv4PHM and SEOBNRv5PHM) has at one or more event where conclusions derived using those two
waveforms differ by more than our ad hoc JS threshold.   These two bottom panels present qualitatively similar results
as previous work \cite{gwastro-mergers-TousifGWTC3} -- see their Figures 3 and 4 --  albeit now performed exclusively with RIFT and using a different set of
waveform models for comparison (i.e., including  SEOBNRv5PHM and IMRPhenomPv2, but omitting IMRPhenomXPHM).

\subsection{Detailed posterior comparisons: Best available models}
Using the  summary statistics provided above and in previous work \cite{gwastro-mergers-TousifGWTC3}, we have identified 
five events that merit further discussion: 
GW200129,
GW191109,
GW200216,
and both events on the date GW200220.  
In an appendix and associated data release, we present  posterior inferences for all events in our sample.   In both the
appendix and our own figures, for comparison and to normalize the scale of differences presented in this work, we also
present the IMRPhenomXPHM results produced for GWTC-3 \cite{LIGO-O3-O3b-catalog}. 
We also present NRSur7dq4 results
provided by a previous study, for validation \cite{gwastro-mergers-TousifGWTC3}.

\begin{figure*}[htbp] 
		\includegraphics[width=0.45\textwidth]{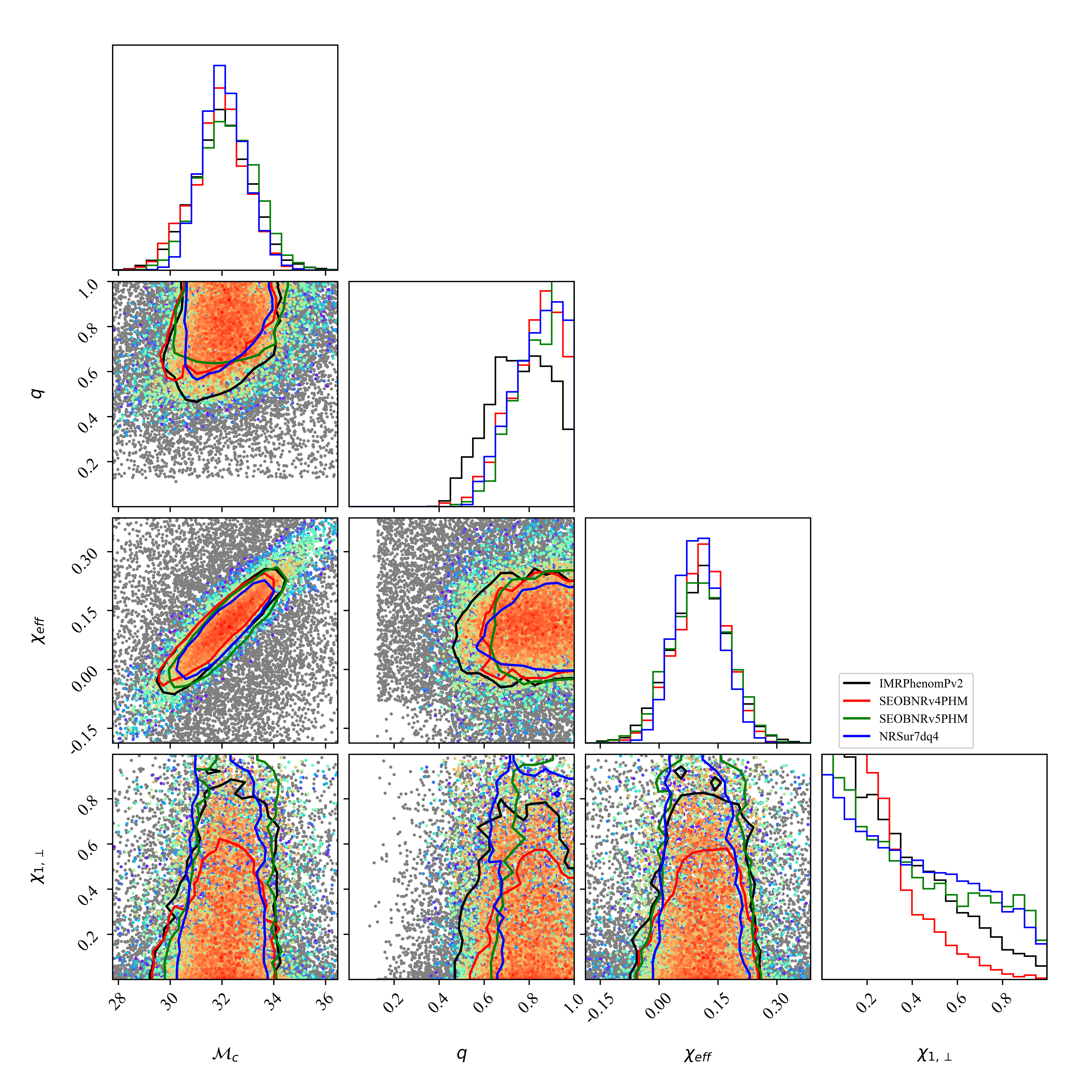}
		\includegraphics[width=0.45\textwidth]{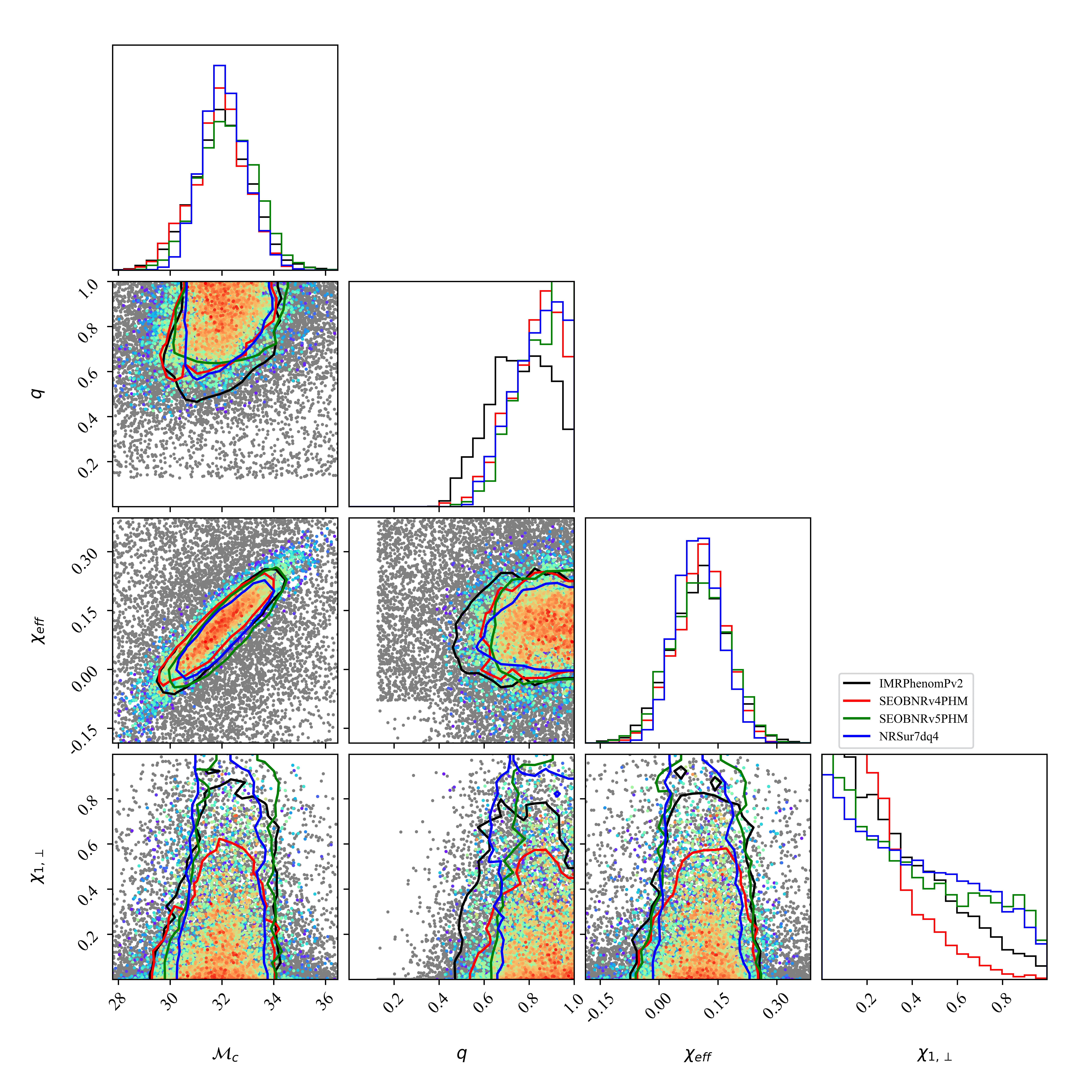}
		\includegraphics[width=0.45\textwidth]{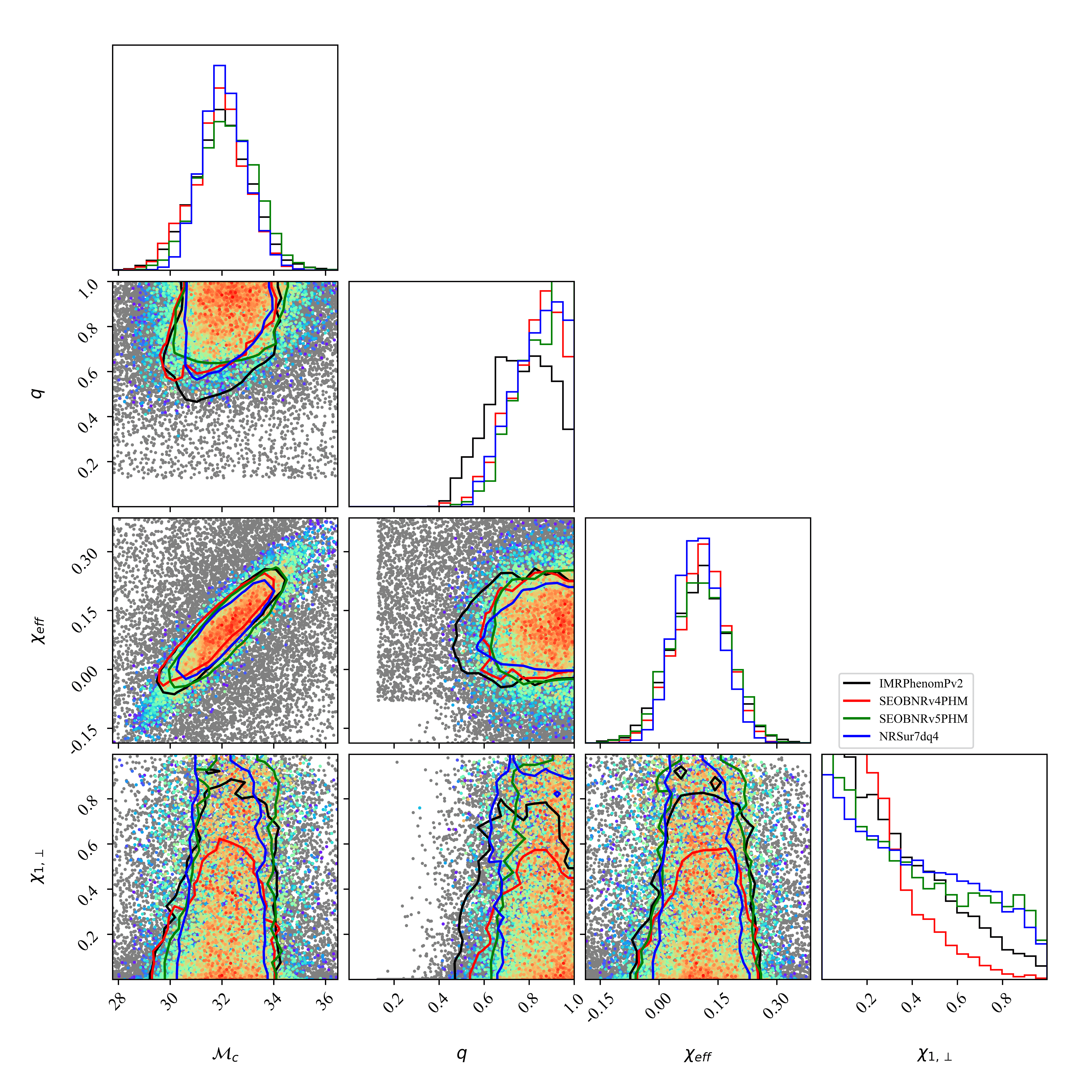}
		\includegraphics[width=0.45\textwidth]{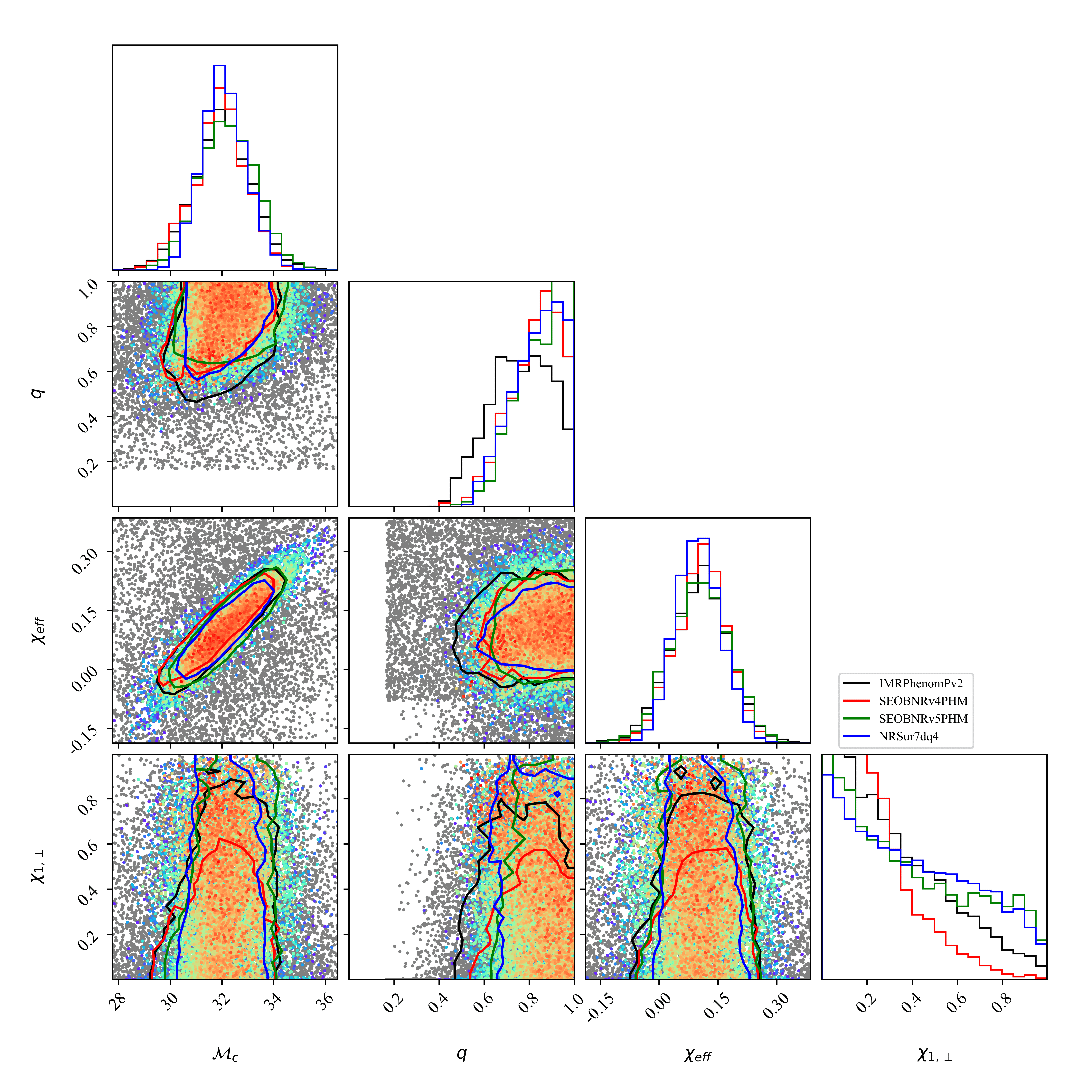}
	\caption{Comparison of waveform models for GW200129\_065458 where the colored data points in the top left panel are from IMRPhenomPv2, the top right panel from SEOBNRv4PHM, the bottom left panel from SEOBNRv5PHM, and the bottom right from NRSur7dq4}
	\label{fig:GW200129_065458} 
\end{figure*}

GW200129 has inspired considerable followup investigation, including claims of precession \cite{2022Natur.610..652H},
eccentricity \cite{2024arXiv240414286G}, and validation studies to assess stability of results under different data cleaning
methods \cite{2022PhRvD.106j4017P,2024PhRvD.109f2006M}. As illustrated in Figure \ref{fig:GW200129_065458}, in our
benchmark analyses, using public cleaned frames and with a (common) independently
generated PSD, we find our three waveform models draw dramatically different conclusions about this event, most notably
the spin magnitudes of the two masses.
For this event, even the two similar models SEOBNRv4PHM and SEOBNRv5PHM lead to notably different conclusions.
Our analysis with NRSur7dq4 favors even more extreme mass ratios, primary spins, and $\chi_p$ than our two SEOBNR results.
However, as discussed below (``comparison to previous results''), our analyses with NRSur7dq4 do not produce as extreme
conclusions about the mass ratio and spin as found in another independent reanalysis \cite{gwastro-mergers-TousifGWTC3},
which employs slightly different settings (e.g., an independently-generated PSD).

\begin{figure}[h]%
	\includegraphics[width=\columnwidth]{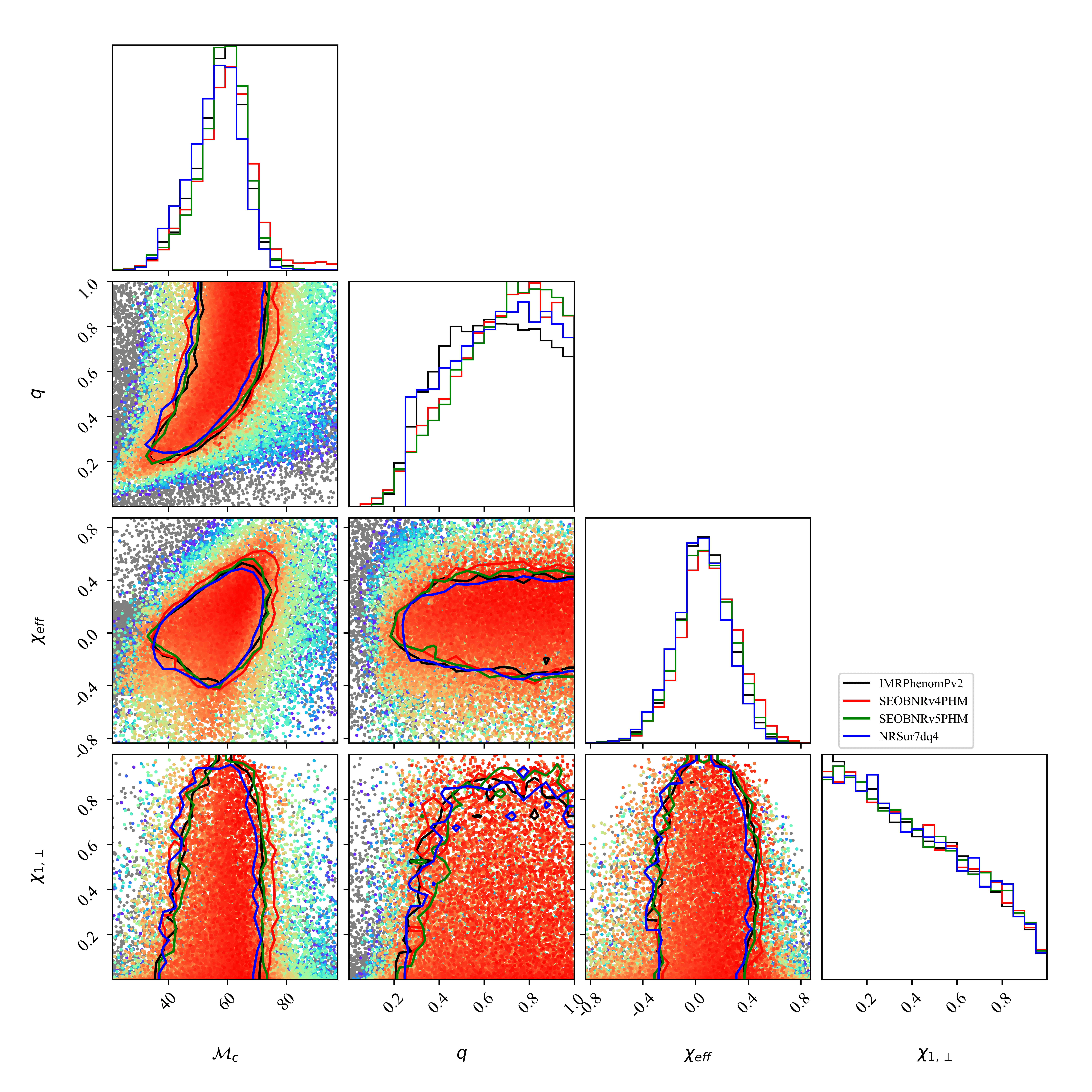}
	\caption{Comparison of waveform models for GW200216\_220804 where the colored data points are approximately the same between IMRPhenomPv2, SEOBNRv4PHM, SEOBNRv5PHM, and NRSur7dq4}
	\label{fig:GW200216_220804}
\end{figure}

\begin{figure}[p]%
	\includegraphics[width=\columnwidth]{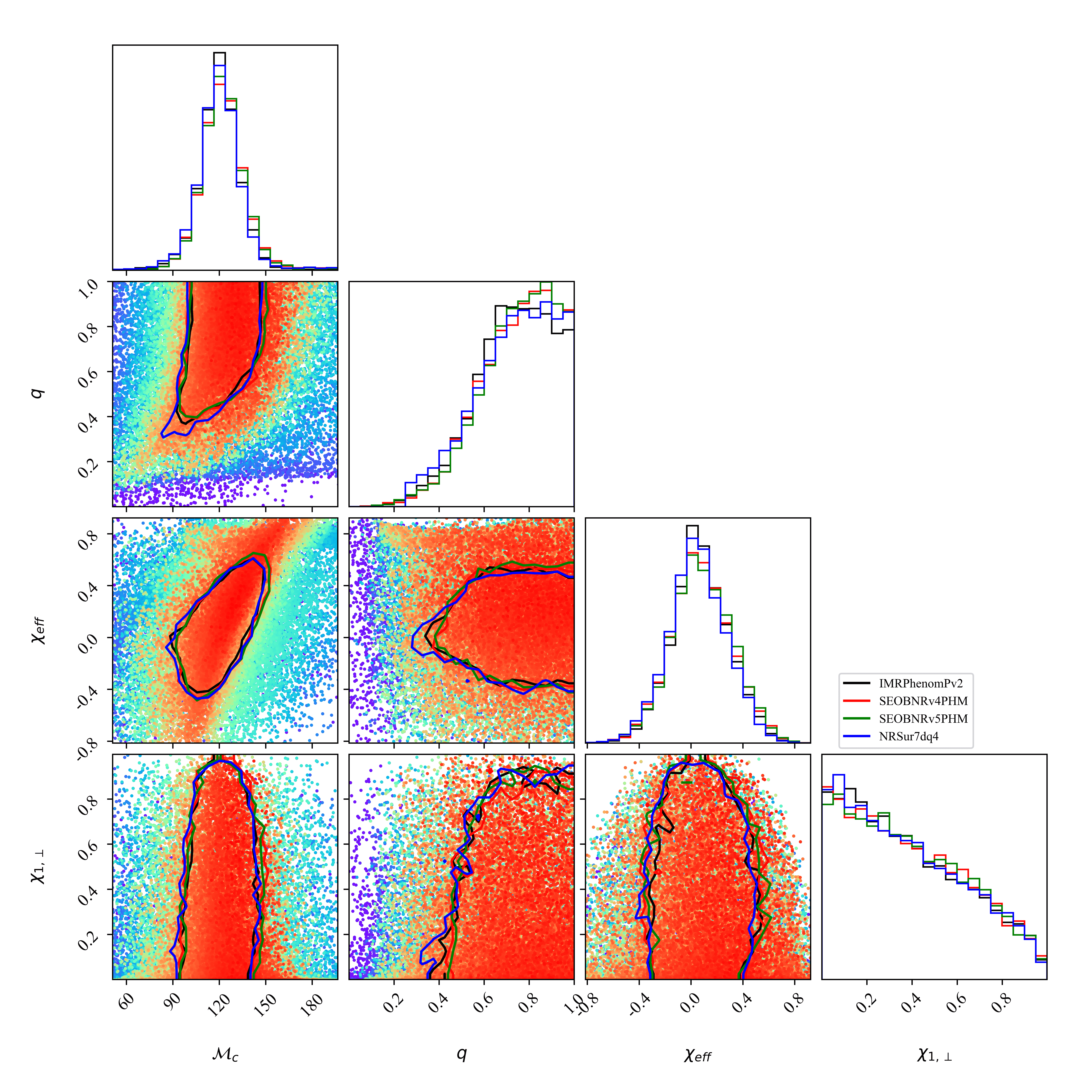}
	\caption{Comparison of waveform models for GW200220\_061928 where the colored data points are approximately the same between IMRPhenomPv2, SEOBNRv4PHM, SEOBNRv5PHM, and NRSur7dq4}
	\label{fig:GW200220_061928}
\end{figure}

\begin{figure}[p]%
	\includegraphics[width=\columnwidth]{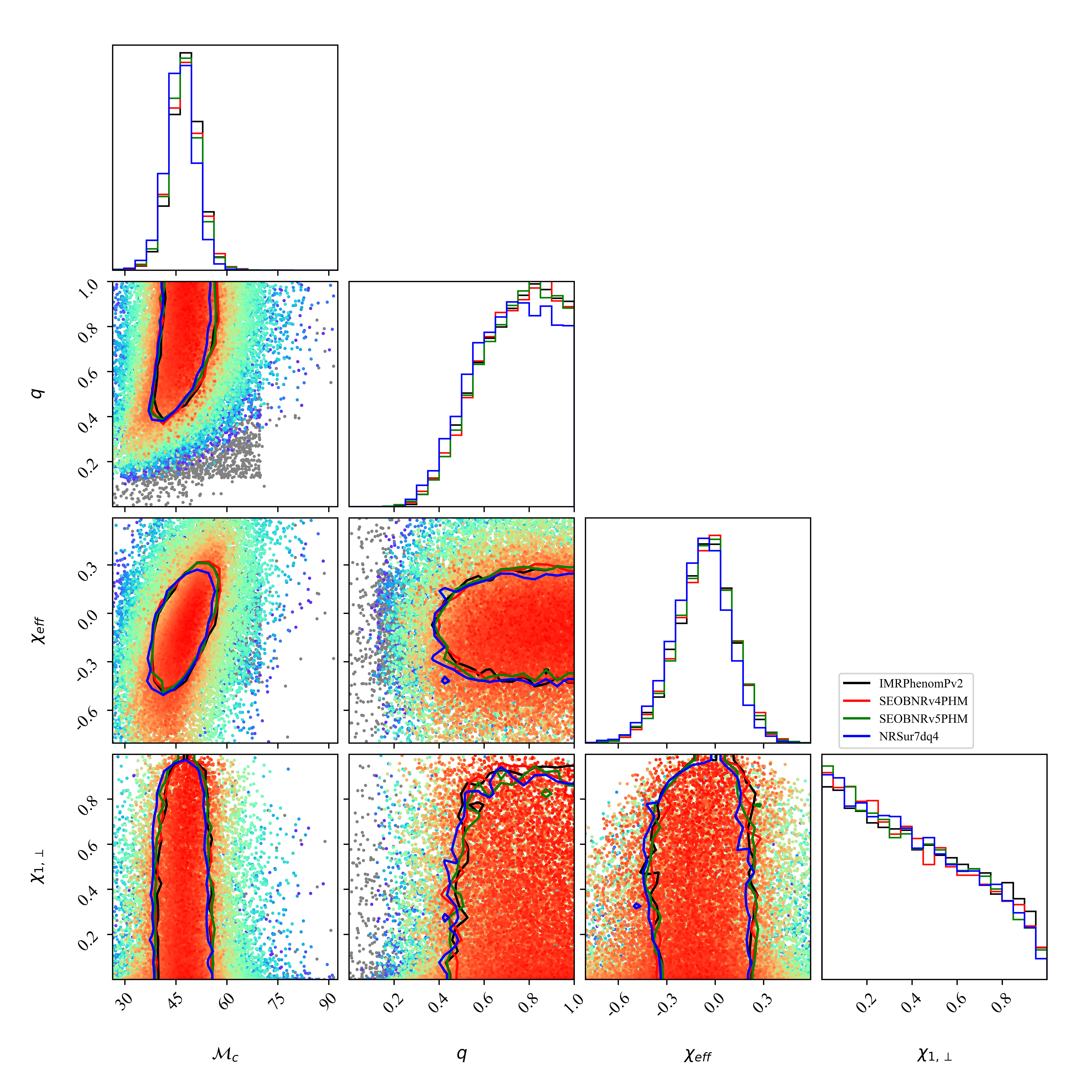}
	\caption{Comparison of waveform models for GW200220\_124850 where the colored data points are approximately the same between IMRPhenomPv2, SEOBNRv4PHM, SEOBNRv5PHM, and NRSur7dq4}
	\label{fig:GW200220_124850}
\end{figure}

By contrast, for  GW200216 (Figure \ref{fig:GW200216_220804}), GW200220\_061928 (Figure \ref{fig:GW200220_061928}), and GW200220\_124850 (Figure \ref{fig:GW200220_124850}), our inferences using the two SEOBNR models are very
consistent with each other, and differ from NRSur7dq4 primarily in the mass ratio and spin magnitude distributions.   In
all three cases, the NRSur7qd4 analysis favors a wider mass ratio distribution and slightly prefers larger spins, while
our two SEOBNR analyses favor comparable mass ratios and lower spins.

\begin{figure*}[htbp] 
		\includegraphics[width=0.45\textwidth]{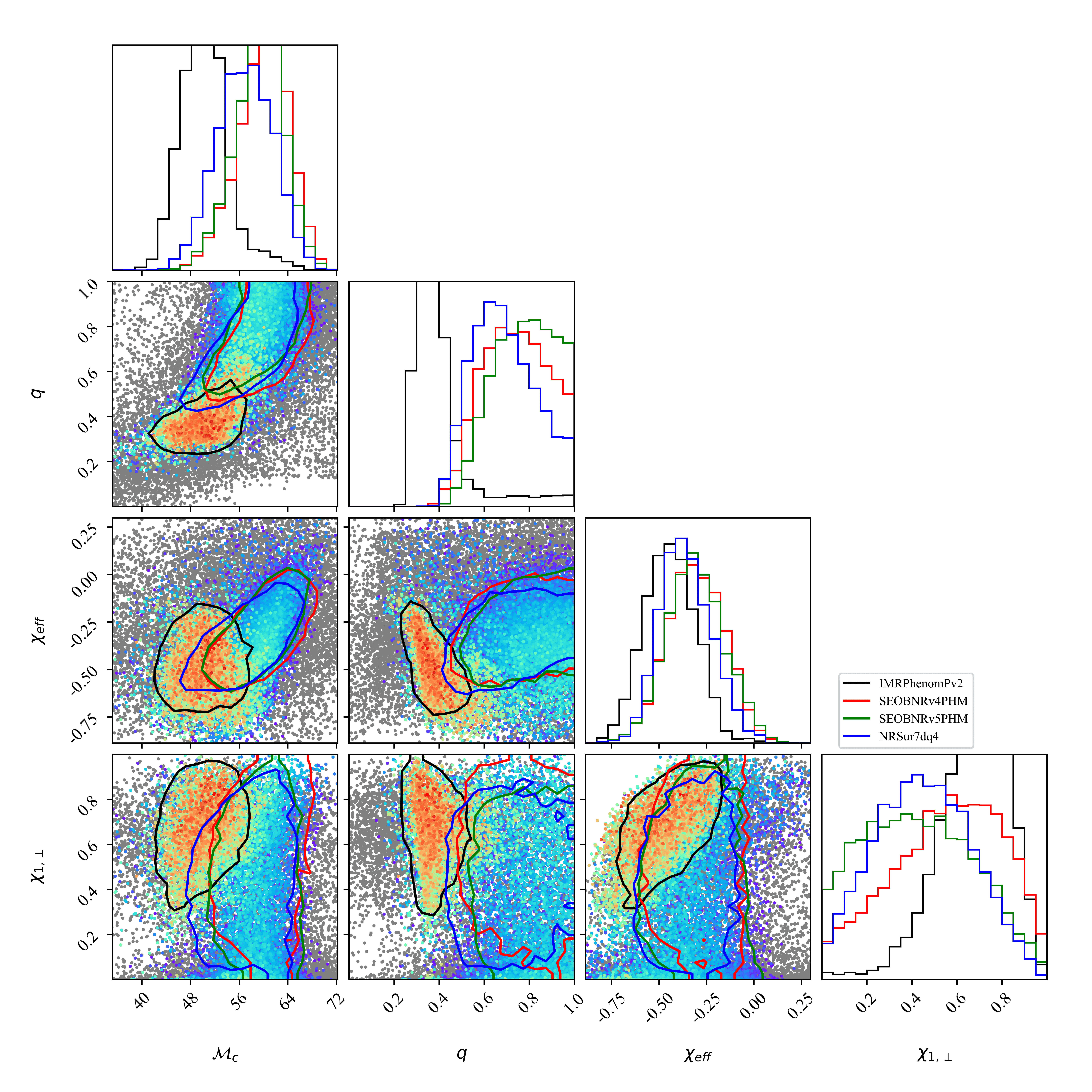}
		\includegraphics[width=0.45\textwidth]{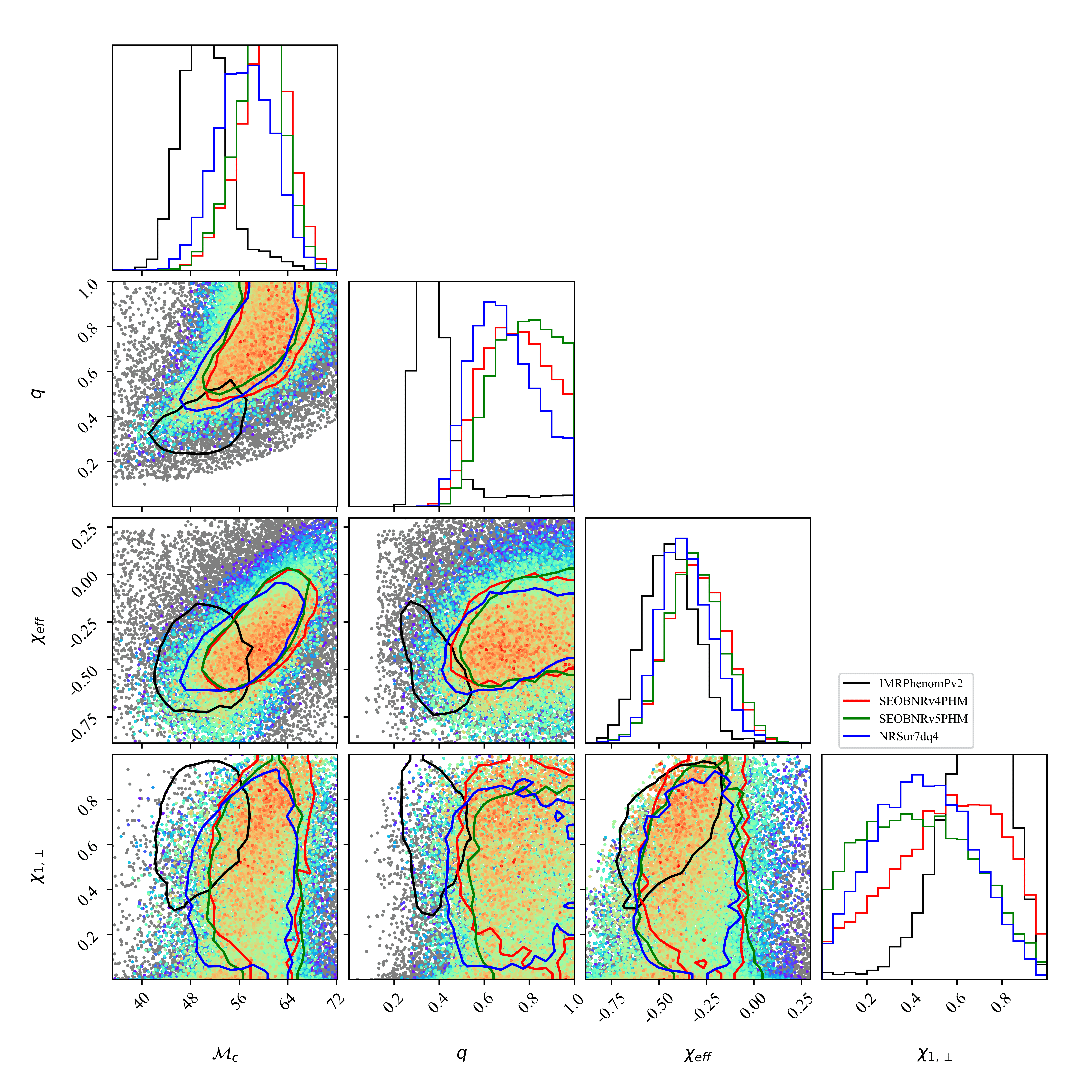}
		\includegraphics[width=0.45\textwidth]{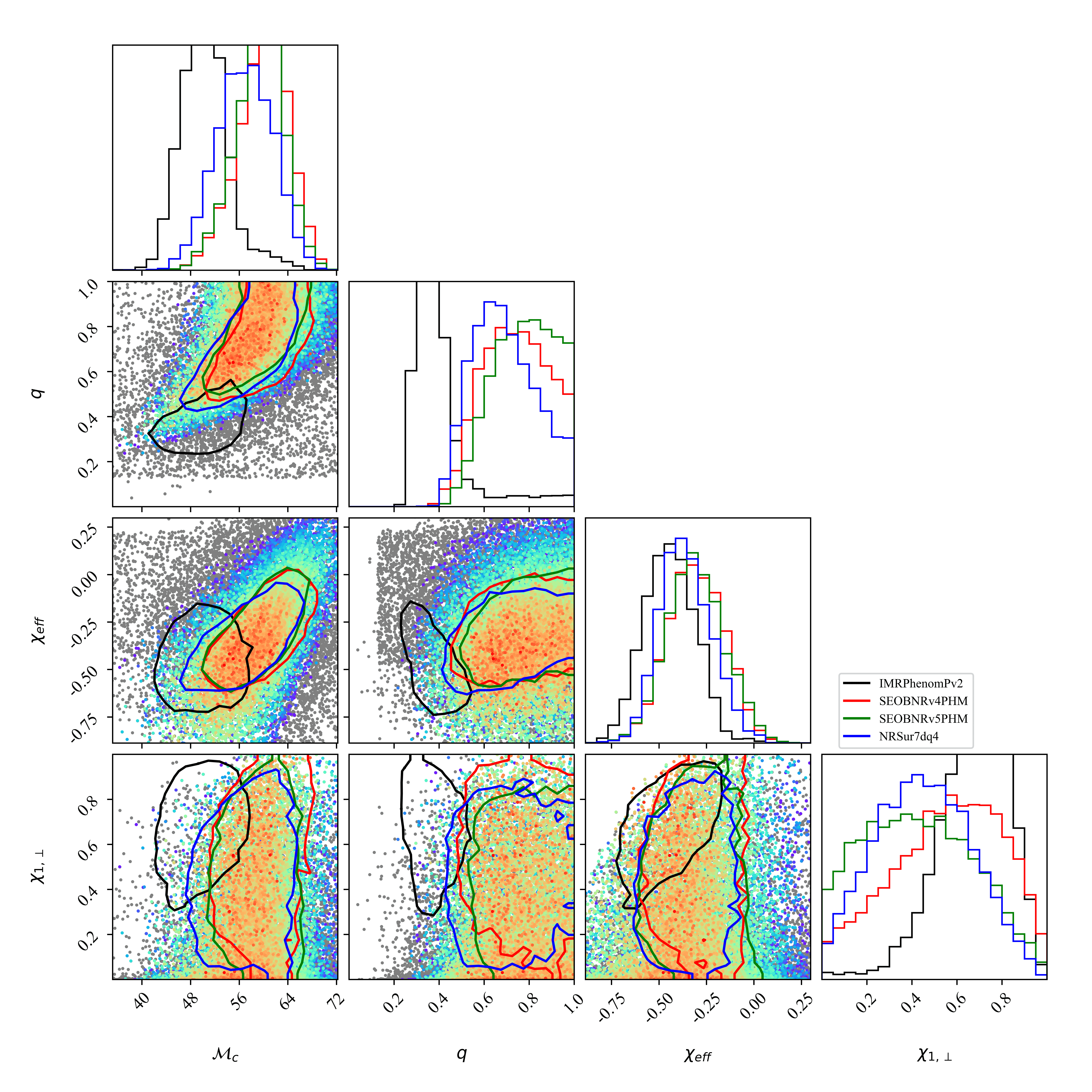}
		\includegraphics[width=0.45\textwidth]{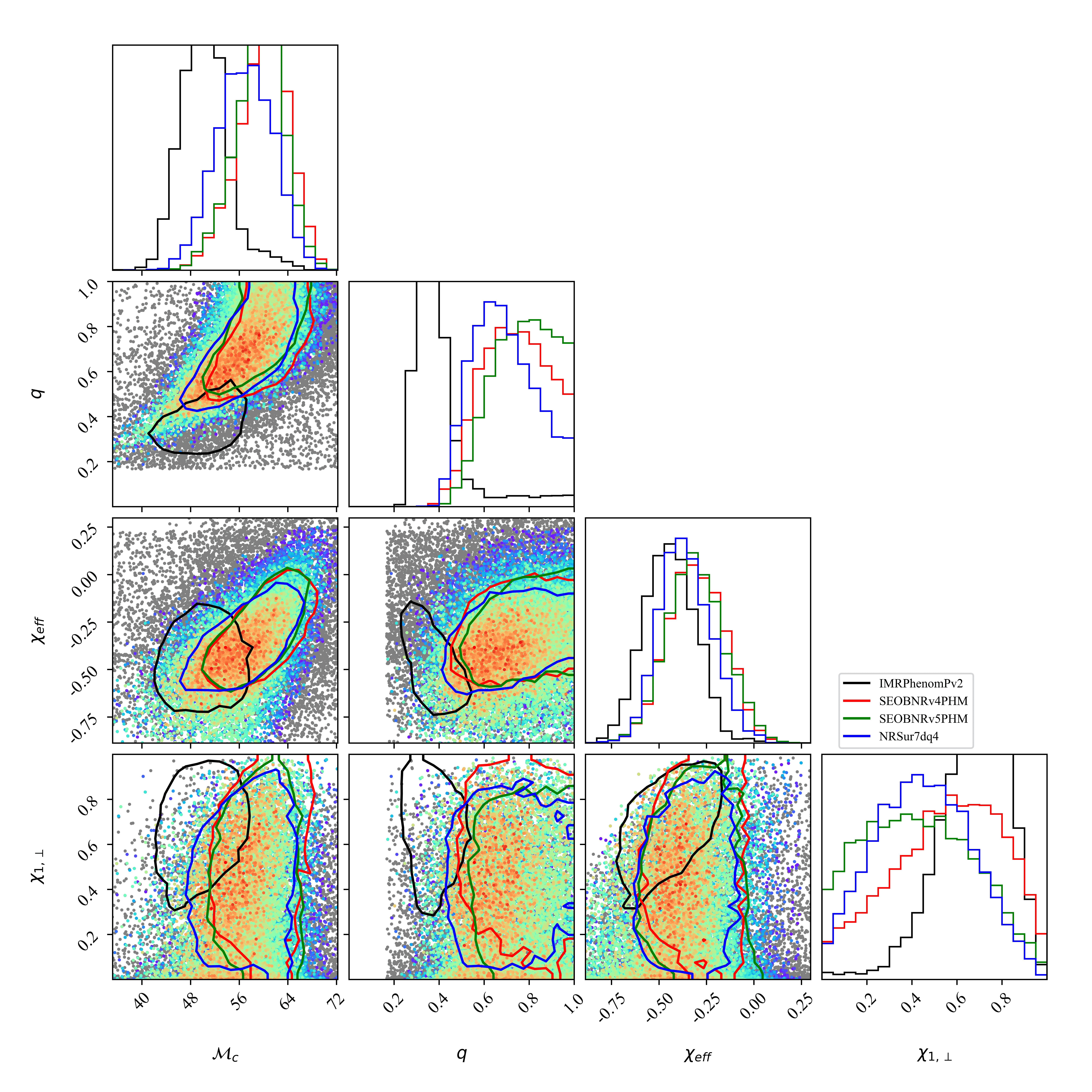}
	\caption{Comparison of waveform models for GW191109\_010717 where the colored data points in the top left panel are from IMRPhenomPv2, the top right panel from SEOBNRv4PHM, the bottom left panel from SEOBNRv5PHM, and the bottom right from NRSur7dq4}
	\label{fig:GW191109_010717} 
\end{figure*}

For GW191109, our inferences once again depend more sentitively on the assumed waveform, as illustrated  in Figures
\ref{fig:GW191109_010717} and \ref{fig:post_1}.  As might be expected given the high mass of the event and the model's
simplified physics, interpretations drawn using  IMRPhenomPv2 in particular produce strikingly different conclusions
about the mass ratio, primary spin, and even $\chi_p$ than analyses performed with any other waveform.   By contrast,
though waveform systematics change the posterior distributions modestly, when ,
comparing analyses performed in this work with state-of-the-art waveforms (SEOBNRv4PHM, SEOBNRv5PHM, and NRSur7dq4), we
find coarsely good overall agreement.

\subsection{Detailed posterior comparisons: IMRPhenomPv2}
Despite the substantial and frequent differences seen between different calculations that both use state-of-the-art waveforms, analyses using IMRPhenomPv2 seem
to be surprisingly comparable to the results derived using all other waveforms in this study;  see the top panels of
Figures \ref{fig:js_batch1} and \ref{fig:js_batch2}.
To investigate this numerical incongruity, in Figures \ref{fig:post_1} and \ref{fig:post_2} we illustrate the posteriors derived using all models, including
IMRPhenomPv2.
For many events and parameters, all one-dimensional posteriors are in good agreement, modulo truncation effects
associated with the finite mass ratio range required for NRSur7dq4.  However, more frequently than for
other waveforms, the posteriors for IMRPhenomPv2 may differ substantially from the consensus conclusions derived with
state-of-the-art waveforms.

\subsection{Other Comparisons with previous work}
 Figures \ref{fig:post_1} and \ref{fig:post_2}  show that for most of the events studied here, our estimates with
 NRSur7dq4, SEOBNRv4PHM, and SEOBNRv5PHM largely agree with results derived using previous analyses with NRSur7dq4 and
 IMRPhenomXPHM.   This agreement is reassuring, given our analysis settings (e.g., the PSD) do not necessarily perfectly
 match the assumptions adopted in these previous works.   However, for two events, these figures reveal  striking
 differences between our analyses and these previous works.

For GW200129, the differences between previous studies and this work are frequent and significant.  Interpreted using the assumptions
of the previous study, both IMRPhenomXPHM and NRSurd7q4 would suggest a posterior extending to mass ratio down to and
peaking near $0.4$, and primary spins that are preferentially very large.  By contrast, our analyses generally prefer
nearly comparable mass ratios and show no preference towards large primary spin; indeed, all our analyses point against
primary spin.  To validate the robustness of our conclusions, Figure \ref{fig:GW200129_065458} shows the underlying marginal  likelhood data used to build our
posterior distributions.  In all cases, we don't find points with high marginal likelihood at very asymmetric mass ratio
$q<0.4$, nor (to a lesser extent) with near-extremal transverse spin.

For GW191109, the differences between previous studies and this work are less extreme but still noteworthy, with both
external posterior results (IMRPhenomXPHM and NRSur7dq4) exhibiting unexpected structure relative to the consensus conclusions
presented in this work.    For example, the external NRSur7dq4 posterior favors a different total mass and strongly
disfavors equal mass.   The external IMRPhenomXPHM posterior favors a wide mass distribution and unexpected complexity
in its conclusions about $\chi_{\rm eff}$ and mass ratio.   By contrast, using the assumptions employed in this work,
our different approximations produce qualitatively similar results, though differing in details (e.g., as noted for this
event SEOBNRv4PHM and SEOBNRv5PHM ddraw slightly different conclusions).

For two more events (GW200220\_124850 and GW200216) we see more modest differences between the results in this work and
our reference analysis using 
NRSur7dq4.   For other events (e.g., GW200224 and GW200220\_061928), the external analysis performed with IMRPhenomXPHM
seems to be an outlier relative to the consensus provided by other analyses, including the external NRSur7dq4 result.
We point out these differences to help target  future investigations such as head-to-head replication studies with
different inference codes, which is outside the scope of this work.

\section{Conclusions}

In this work, we demonstrate reanalysis of several O3 binary black hole mergers, using multiple state-of-the-art
waveform models with consistent analysis settings.   Combining the \texttt{asimov} production-quality
inference automation and the \texttt{RIFT} distributed parameter inference code, our proof-of-concept calculations
show concretely how to perform these and related  analyses at scale, 
without requiring access to special-purpose or closely-held computing or data resources.    
Our  analysis is impactful because of the systematic differences this
specific study illuminates: in astrophysically tantalizing real events, binary black hole
parameter inferences can draw very different conclusions about source parameters.    This framework and related studies
like it remain a critical ingredient to better understand how precisely and accurately we can infer source (and thus
population) parameters of gravitational wave sources, given our still-imperfect understanding of how to approximate
the gravitational waves emitted from  compact binary mergers.

In this work, we have emphasized real high-mass sources primarily due to their high detection rate: more  sources
provides more opportunity to identify significant systematic differences between different models.   That said, we
anticipate that the primary utility of the specific approach outlined in this work will be highest for longer sources,
whose many cycles tightly constrain binary parameters  and which could better illuminate systematic differences, as in
GW190412.   
Too, we anticipate that for the next few years at least, ongoing waveform development will only increase the number and kind of waveform comparison
studies needed for each event.  For example, even in this study, we have neglected both physical degrees of freedom
(e.g., eccentricity) and alternative waveform families (e.g., TEOBResumS).

\section{Acknowledgments}
This material is based upon work supported by NSF's LIGO Laboratory which is a major facility fully funded by the National Science Foundation.

\appendix
\section{Posteriors for all events}
\begin{figure*}
	\includegraphics[width=\textwidth]{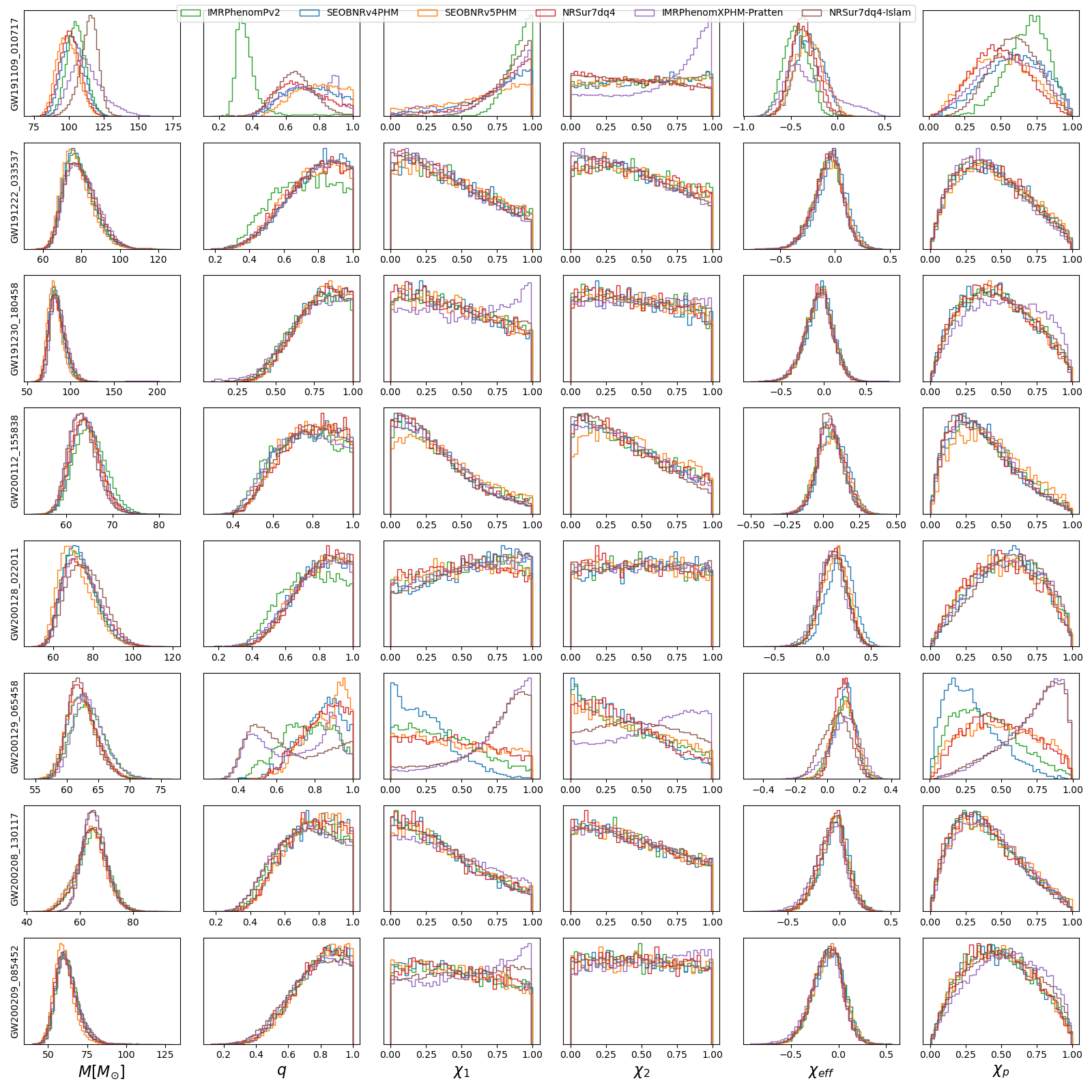}
	\caption{Posteriors for the source-frame total mass $M$, mass ratio $q$, spin magnitudes $\chi_1$ and $\chi_2$, effective inspiral spin parameter $\chi_{eff}$, and spin precession parameter $\chi_p$ for selected events that have the most significant differences between results obtained using IMRPhenomPv2 (green), SEOBNRv4PHM (blue), SEOBNRv5PHM (orange), NRSur7dq4 (red), IMRPhenomXPHM (purple) and NRSur7dq4-Islam (brown). Here, "NRSur7dq4-Islam" is the plotted NRSur7dq4 results obtained from \cite{gwastro-mergers-TousifGWTC3}, and IMRPhenomXPHM from \cite{gwastro-mergers-IMRPhenomXP}}.
	\label{fig:post_1}
\end{figure*}

\begin{figure*}
  	\includegraphics[width=\textwidth]{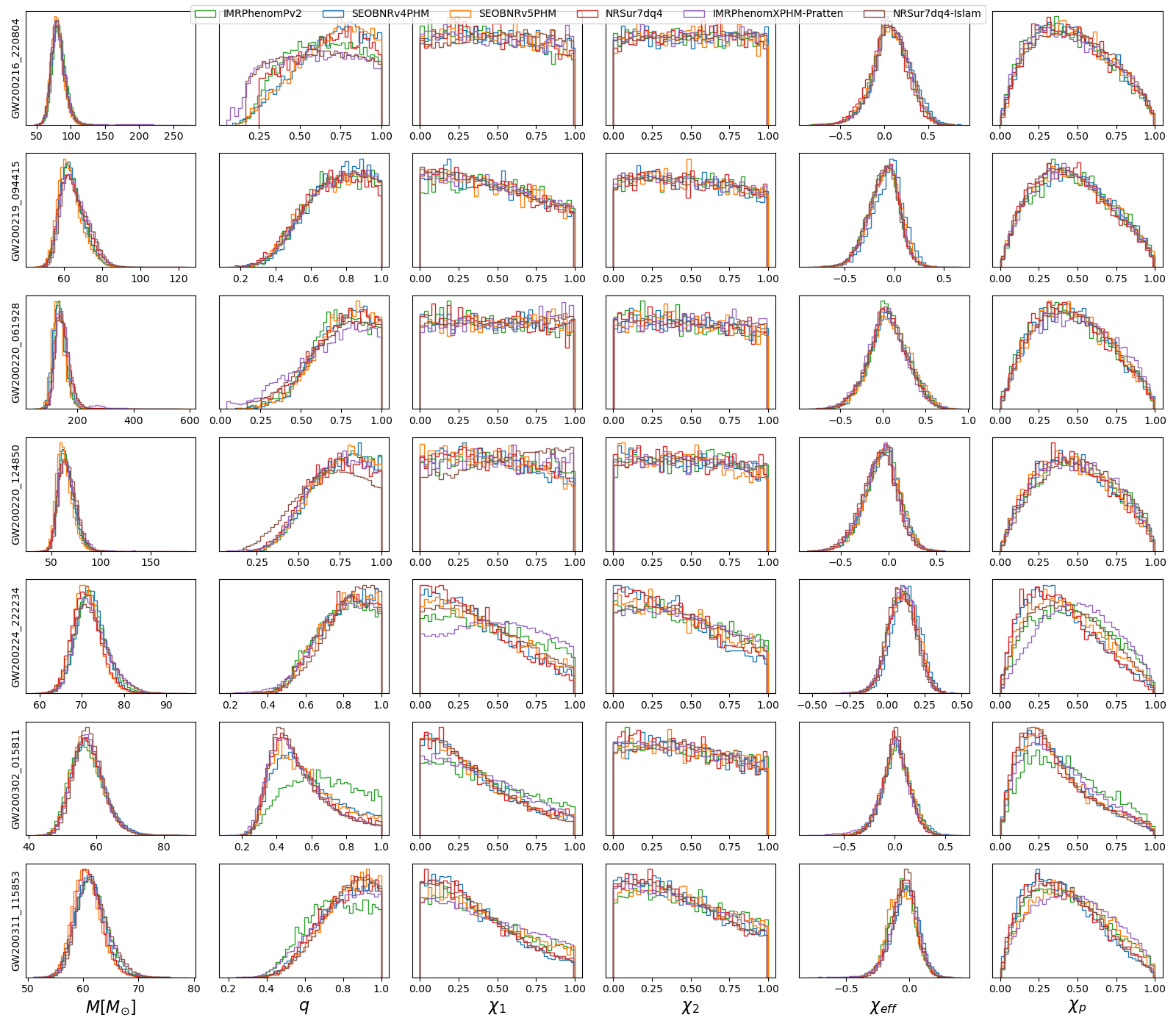}
	\caption{Posteriors for the source-frame total mass $M$, mass ratio $q$, spin magnitudes $\chi_1$ and $\chi_2$, effective inspiral spin parameter $\chi_{eff}$, and spin precession parameter $\chi_p$ for selected events that have the most significant differences between results obtained using IMRPhenomPv2 (green), SEOBNRv4PHM (blue), SEOBNRv5PHM (orange), NRSur7dq4 (red), IMRPhenomXPHM (purple) and NRSur7dq4-Islam (brown). Here, "NRSur7dq4-Islam" is the plotted NRSur7dq4 results obtained from \cite{gwastro-mergers-TousifGWTC3}, and IMRPhenomXPHM from \cite{gwastro-mergers-IMRPhenomXP}}.
	\label{fig:post_2}
\end{figure*}

Figure \ref{fig:post_1} shows one-dimensional marginal posteriors for the first eight events presented in this study, and  Figure \ref{fig:post_2} shows the other seven. In these
figures, the IMRPhemonPv2, SEOBNRv4PHM, SEOBNRv5PHM and NRSur7dq4 results were generated as a part of our investigation.  By contrast, the
NRSur7dq4-previous results are provided for comparison and were previously reported in \cite{gwastro-mergers-TousifGWTC3}, and IMRPhenomXPHM were previously reported in \cite{gwastro-mergers-IMRPhenomXP}.

\clearpage

\bibliography{LIGO-publications,gw-astronomy-mergers-approximations,gw-astronomy-mergers,references,gw-astronomy-detection}

\end{document}